# A unified theoretical framework for fluctuating-charge models in atom-space and in bond-space

Jiahao Chen and Todd J. Martínez

Department of Chemistry, Frederick Seitz Materials Research Laboratory and

The Beckman Institute

University of Illinois at Urbana Champaign

Urbana, IL 61801

**Abstract**

Our previously introduced QTPIE (charge transfer with polarization current equilibration) model (J. Chen and T. J. Martínez, Chem. Phys. Lett. **438**, 315 (2007)) is a fluctuating-charge model with correct asymptotic behavior. Unlike most other fluctuating-charge models, QTPIE is formulated in terms of charge-transfer variables and pairwise electronegativities, not atomic charge variables and electronegativities. The pairwise character of the electronegativities in QTPIE avoids spurious charge transfer when bonds are broken. However, the increased number of variables leads to considerable computational expense and a rank-deficient set of working equations, which is numerically inconvenient. Here, we show that QTPIE can be exactly reformulated in terms of atomic charge variables, leading to a considerable reduction in computational complexity. The transformation between atomic and bond variables is generally applicable to arbitrary fluctuating charge models, and uncovers an underlying topological framework that can be used to understand the relation between fluctuating-charge models and the classical theory of electrical circuits.

# Introduction

Fluctuating-charge models, also known as electronegativity equalization models or chemical potential equilibration models, are a computationally efficient way of incorporating polarization and charge-transfer effects into classical molecular dynamics simulations.[1-6] In fluctuating-charge models, the response of a molecular charge distribution to electrostatic interactions is represented by a redistribution of atomic charges within the system, driven by atomic electronegativities and hardnesses. Polarization and charge transfer are therefore modeled as different aspects of the same phenomenon, the only difference being whether the charge transfer occurs within or without any given molecule. The first fluctuating-charge models to be incorporated into dynamical simulations are the well-known QEq[7-11] and fluc-q[12-16] methods, although related methods had been introduced even earlier to determine atomic charges at equilibrium geometries.[17-25]

Fluctuating-charge models are computationally tractable even for large molecules, increasing the expense by less than an order of magnitude compared to molecular dynamics with fixed-charge force fields. However, existing models are known to overestimate the amount of charge transfer occurring when bonds are highly stretched, leading to nonzero charge transfer at infinite separation even for isolated molecules.[26-31] To address this problem, we introduced the QTPIE charge model[27,29] with an energy function of the form

$$E = \sum_{i=1}^{N}\sum_{j=1}^{N} p_{ji}f_{ij} + \tfrac{1}{2}\sum_{i=1}^{N}\sum_{j=1}^{N}\sum_{k=1}^{N}\sum_{l=1}^{N} p_{ki}p_{lj}J_{ij} \qquad (1)$$

The indices $i$, $j$, $k$ and $l$ in Eq. (1) run over the $N$ atoms that constitute the fluctuating charge sites. Charge transfer variables accounting for the amount of charge transferred from atom $j$ to atom $i$ are denoted as $p_{ji}$. Symmetry considerations dictate that the charge transfer variables be antisymmetric, i.e.

$$p_{ji} = -p_{ij} \qquad (2)$$

Recent related work by Nistor and coworkers[28] also makes use of these charge transfer variables, which they call split-charge variables. The atomic electronegativities originally used in QEq and other fluctuating charge models are generalized to explicit pairwise electronegativities,[26,27] $f_{ij}$. Finally, $J_{ij}$ is the hardness matrix with atomic chemical hardnesses on the diagonal entries and screened Coulomb interactions on the off-diagonal entries.

The values of the charge-transfer variables are determined by minimizing the energy expression in Eq. (1). The antisymmetry of the charge transfer variables implies that the equations can be simplified by introducing a single pair index $\lambda(i,j)$ which runs over unique atom pairs:

$$\lambda(i,j) = \tfrac{1}{2}\max(i,j)\big(\max(i,j)-1\big) + \min(i,j) \qquad (3)$$

The variables $\{p_\lambda\}$ that minimize Eq. (1) are solutions to the following linear system of equations

$$\begin{aligned}
\frac{\partial E}{\partial p_\lambda} &= \sum_\nu A_{\lambda\nu} p_\nu - V_\lambda = 0 \\
V_{\lambda(i,j)} &= f_{ij} - f_{ji} \\
A_{\lambda(i,j)\nu(k,l)} &= \frac{1}{2}\frac{\partial^2 E}{\partial p_\lambda \partial p_\nu} = \tfrac{1}{2}\big(J_{ik} + J_{jl} - J_{jk} - J_{il}\big)
\end{aligned} \qquad (4)$$

where $V$ is a vector of pairwise electronegativity differences, which can be interpreted as potential differences between atomic pairs. The real and symmetric hardness matrix $\mathbf{A}$ thus represents a linear map from charge transfer variables into voltages. Given the physical meaning of the charge transfer variables, it is clear that the atomic charges can be recovered from the solution to Eq. (4) using the relation

$$q_i = \sum_j p_{ji} \tag{5}$$

As demonstrated previously,[27,29] this model successfully solves the problem of nonvanishing charge-transfer at infinite separation. However, the current formulation is computationally cumbersome for large molecules because Eq. (4) involves solving a matrix problem of dimension $\frac{1}{2}N(N-1) = O(N^2)$, implying that an implementation of the QTPIE model based on naïve direct solvers for Eq. (4) has a computational complexity of $O(N^6)$. This cost can be reduced using iterative methods to $O(N^4)$, and exploiting sparsity could reduce the cost further to $O(N^2)$. However, the prefactor is expected to be large due to the long-range nature of the Coulomb interaction; furthermore, the matrix $\mathbf{A}$ is rank-deficient, which necessitates using more costly numerical algorithms such as singular value decomposition (SVD)[32] to solve the QTPIE equations.

In this work, we investigate the origins of the rank deficiency of $\mathbf{A}$ and present more practical methods to solve the QTPIE model. We show that in addition to SVD, a complete orthogonal factorization technique exists for solving Eq. (4) more efficiently. Furthermore, we illustrate a procedure for discovering a transformation that complements Eq. (5) by relating charge-transfer variables explicitly in terms of atomic charge variables. This allows a completely exact reformulation of the QTPIE working equations explicitly in terms of atomic charges. This reformulated QTPIE model is of the same computational complexity as other fluctuating-charge models such as EEM,[24,25] QEq,[7,8] *fluc*-q,[12-16] AACT,[33-35] CPE,[13,36,37] ABEEM,[38,39] and others[3,17,20,22,30,40-54] that are expressed in terms of atomic charge variables.

# Null modes of the capacitance matrix

For systems with $N > 2$ atoms, the matrix $\mathbf{A}$ in Eq. (4) is rank deficient. As an illustration, we diagonalize $\mathbf{A}$ for a single water molecule in its equilibrium ground-state geometry. The details of the atomic hardnesses and orbitals defining the elements of the matrix $\mathbf{A}$ may be found in our previous work.[29] However, the results shown are general and not dependent on these details. The eigenvectors of $\mathbf{A}$ may be thought of as normal modes for charge flow in the system. These bond-space eigenvectors are visualized in Figure 1 along with the atom-space eigenvectors of the Coulomb matrix $\mathbf{J}$, which is the hardness matrix for QEq. Even for this small triatomic molecule, $\mathbf{A}$ has a non-trivial kernel or nullspace. In this case, it is spanned by the vector $u^{\perp} = \frac{1}{\sqrt{3}}(-1,1,-1)$ that describes cyclic charge transport. The effect of the kernel is given by the scalar product

$$u^{\perp} \cdot p = 0 = \sum_{\lambda} u_{\lambda}^{\perp} p_{\lambda} = \frac{1}{\sqrt{3}}\left(-p_{21} + p_{31} - p_{32}\right) = \frac{1}{\sqrt{3}}\left(p_{12} + p_{23} + p_{31}\right) \tag{6}$$

showing that this combination of the charge-transfer variables cannot contribute any net potential difference to the system. This is closely related to Kirchhoff's voltage law, i.e. there is no change in the electrostatic potential when a charge is transported about a closed loop. This law reflects the conservative nature of the electrostatic potential embodied by the matrix $\mathbf{A}$.

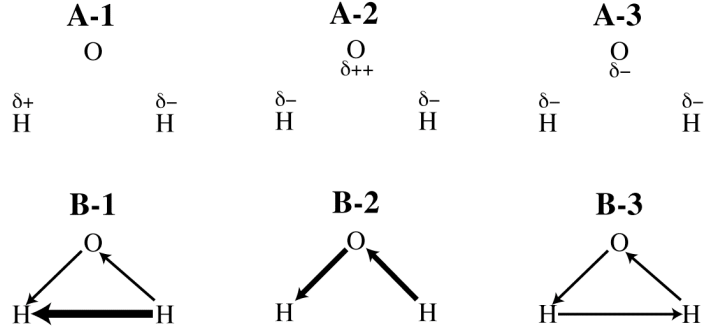

Figure 1. Visual representations of eigenvectors of the atom space hardness matrix $\mathbf{J}$ of Eq. (1) (A-1—A-3) and the bond space hardness matrix $\mathbf{A}$ of Eq. (4) (B-1—B-3) for a single water molecule. The δ+, δ- symbols represent increases and decreases in charge on the respective atoms, while arrows show the direction of charge transfer with relative magnitudes indicated by their thicknesses. The respective eigenvalues are 0.181 (A-1), 0.101 (A-2), 1.231 (A-3), 1.273 (B-1), 0.350 (B-2), and 0.000 (B-3). Although the magnitude of the nonzero eigenvalues depends on the detailed parameters used to construct $\mathbf{J}$ and $\mathbf{A}$, the zero eigenvalue corresponding to the eigenvector shown in B-3 does not.

*The physical significance of rank deficiency*

The rank deficiency of $\mathbf{A}$ for $N > 2$ is not an unfortunate numerical accident, but rather is an unavoidable consequence of electrostatics combined with our choice of representation. The second term in Eq. (1) can be rewritten purely in terms of atomic charges as

$$\frac{1}{2}\sum_{i=1}^{N}\sum_{j=1}^{N}\sum_{k=1}^{N}\sum_{l=1}^{N}p_{ki}p_{lj}J_{ij} = \frac{1}{2}\sum_{i=1}^{N}\sum_{j=1}^{N}q_iq_jJ_{ij}, \qquad (7)$$

or in matrix notation, $\frac{1}{2}\vec{p}^T\mathbf{A}\vec{p} = \frac{1}{2}\vec{q}^T\mathbf{J}\vec{q}$. Eq. (5) clearly shows that there exists a linear transformation $\mathbf{T}$ that maps from $\mathbf{p}$ to $\mathbf{q}$ by acting on the left, i.e. $\vec{q} = \mathbf{T}\vec{p}$. Hence there must be a corresponding linear transformation $\mathbf{T}'$ that maps from $\mathbf{q}$ to $\mathbf{p}$ by acting on the right, i.e. $\vec{q}^T = \vec{p}^T\mathbf{T}'$. We will later state $\mathbf{T}$ explicitly in Eqs. (13) and (14), and construct $\mathbf{T}'$ in Eqs. (15) and (16); however, their existence is sufficient to show that by associativity,

$$\vec{p}^T \mathbf{A} \vec{p} = \vec{q}^T \mathbf{J} \vec{q} = \left( \vec{p}^T \mathbf{T}' \right) \mathbf{J} \left( \mathbf{T} \vec{p} \right) = \vec{p}^T \left( \mathbf{T}' \mathbf{J} \mathbf{T} \right) \vec{p} \tag{8}$$

This shows that $\mathbf{A}$ and $\mathbf{J}$ are related by a linear transformation, i.e. $\mathbf{A} = \mathbf{T}' \mathbf{J} \mathbf{T}$, and that the ranks of $\mathbf{A}$ and $\mathbf{J}$ are equal since the transformation is an information-preserving projection from atom space into bond space. Therefore both $\mathbf{A}$ and $\mathbf{J}$ are of rank $N$, and the problem has rank $N - 1$ since there is one constraint of electrical neutrality which is implicitly enforced by the skew–symmetry of the charge–transfer variables. Therefore, there are only $N - 1$ physically significant degrees of freedom.

We can also interpret the rank of $\mathbf{A}$ as a consequence of Kirchhoff's voltage law. We introduce a graph $G = (V, E)$ as a convenient bookkeeping construct, with each vertex $v \in V$ corresponding to an atom or its charge, and each edge $e \in E$ corresponding to a unique charge-transfer variable $p_\lambda$. An arbitrary set of charge–transfer variables can then be mapped onto a corresponding graph $G$. The relevant physics is expressed by Kirchhoff's voltage law, which states that the change in potential as charge is transported about a closed loop vanishes. Therefore, every set of charge-transfer variables that map onto a graph containing a closed loop is linearly dependent. Hence linearly independent sets of charge–transfer variables that must correspond to graphs $G$ that do not contain cycles. At the same time, charges are allowed to flow between any pair of atoms in our model, unlike other models that enforce *a priori* constraints on pairwise charge flow.[3,5,12-15,20,48-51,55,56] Hence by definition, the physically interesting sets of charge–transfer variables must have graphs $G$ that are spanning trees. An elementary result of graph theory immediately yields that trees that connect all $N$ vertices of $G$ have $N - 1$ edges, since adding any more edges would introduce a cycle. Hence in order for QTPIE to be consistent with the conservative nature of the electrostatic potential, only $N - 1$ charge–transfer variables can be linearly independent.

In summary, the linear dependency of the full set of charge-transfer variables is reflected in the rank-deficiency of **A** in Eq. (4). The EPAPS contains a formal mathematical proof using the theory of matroids[57,58] showing the existence of an isomorphism between the combinatorial properties of matrices and the combinatorial properties encapsulated in suitably defined graphs.

*Numerical solution of the rank-deficient system in Eq. (4)*

The singular and indefinite nature of **A** necessitates a careful choice of numerical algorithm to solve the QTPIE equations. Singular value decomposition (SVD) has previously been used in the context of electronegativity equalization methods,[29,42] but it can be computationally expensive. Here, we describe a faster algorithm employing complete orthogonal decomposition[59] (COD) which identifies and projects out the nullspace;[32] this is formally equivalent to the method used to find the minimum-norm least-squares solution for underdetermined equations.

The key transformation that allows the nullspace of a matrix to be identified numerically is the rank-revealing QR factorization,[32] which is an orthogonal factorization that employs column pivoting to separate the range of a matrix from its kernel. Rank-revealing QR decomposes a square matrix **A** of dimension $M$ and rank $\rho$ into the matrix product

$$\mathbf{A} = \mathbf{Q}\begin{pmatrix} \tilde{\mathbf{R}} & \tilde{\mathbf{S}} \\ \mathbf{0} & \mathbf{0} \end{pmatrix}\mathbf{P^{-1}} \tag{9}$$

where **Q** is an orthogonal $M$ x $M$ matrix, $\tilde{\mathbf{R}}$ is an upper triangular $\rho$ x $\rho$ matrix, $\tilde{\mathbf{S}}$ is a rectangular $\rho \times (M - \rho)$ matrix, and **P** is a permutation matrix describing the sequence of pivots used to compute the factorization. Furthermore, it is possible to construct a complete orthogonal decomposition from Eq. (9) of the form

$$\mathbf{Q}^{T}\mathbf{A}\left(\mathbf{P}\mathbf{Q}\right)=\begin{pmatrix}\tilde{\mathbf{R}} & \tilde{\mathbf{S}} \\ \mathbf{0} & \mathbf{0}\end{pmatrix}\mathbf{P}^{-1}\left(\mathbf{P}\mathbf{Q}\right)=\begin{pmatrix}\tilde{\mathbf{T}} & \mathbf{0} \\ \mathbf{0} & \mathbf{0}\end{pmatrix} \qquad (10)$$

Since $\mathbf{A}$ is symmetric, $\mathbf{Q}^{T}$, and therefore its permutation $\mathbf{P}\mathbf{Q}^{T}$, can act from the right to zero out $\tilde{\mathbf{S}}$, projecting its useful information into the square, full-rank submatrix $\mathbf{T}$ of dimension $\rho \, \mathrm{x} \, \rho$. The COD given in Eq. (10) is sufficient to construct an algorithm to solve $\mathbf{A}\vec{p}=\vec{V}$, which can now be written as

$$\mathbf{Q}^{T}\mathbf{A}\vec{p}=\begin{pmatrix}\tilde{\mathbf{T}} & \mathbf{0} \\ \mathbf{0} & \mathbf{0}\end{pmatrix}\left(\mathbf{Q}^{T}\mathbf{P}^{-1}\right)\vec{p}=\mathbf{Q}^{T}\vec{V} \qquad (11)$$

This equation shows explicitly that only the rows of $\mathbf{Q}$ that span the range of $\rho$ contribute to the norm of the solution. It is therefore useful to define a partition of $\mathbf{Q}=\begin{pmatrix}\mathbf{U} & \mathbf{Z}\end{pmatrix}^{T}$, where $\mathbf{U}$ is the rectangular matrix $\mathbf{U}$ of dimension $\rho \times M$ which is formed by the first $\rho$ rows of $\mathbf{Q}$. We can therefore calculate the minimum-norm solution $\vec{p}_{0}$ to (11) using

$$\tilde{\mathbf{T}}\left(\mathbf{U}^{T}\mathbf{P}\vec{p}_{0}\right)=\mathbf{U}^{T}\vec{V} \qquad (12)$$

which can be solved using conventional techniques such as Cholesky factorization for $\mathbf{U}^{T}\vec{p}_{0}$; left multiplication by $\mathbf{P}\mathbf{U}$ completes the algorithm.

We note that had $\tilde{\mathbf{T}}$ been diagonalized, we would have solved the problem using SVD; the computational savings of using this COD algorithm arises precisely from our ability to solve the equations without a complete diagonalization. This results in a reduction in asymptotic complexity from $O\left(M^{3}\right)$ in SVD to $O\left(\rho M^{2}\right)$ in COD.[32]

## Transformations between bond and atom space

When applied to the QTPIE model given in Eq. (4), which has dimension $M = \frac{1}{2}N(N-1) = O(N^2)$ and rank $\rho = N - 1 = O(N)$, the COD algorithm scales as $O(N^5)$ while SVD scales as $O(N^6)$. This therefore represents significant savings in computational costs. However, both algorithms remain costly as the problem is formulated in the space of charge-transfer variables. Perhaps surprisingly, it is possible to derive an explicit transformation from the set of charge-transfer variables to the set of atomic charge variables, thus enabling the reformulation of Eq. (4) in a space of significantly smaller dimensionality.

Again, graph theory provides a framework for understanding why. The transformation of variables arises from a dual relationship between the vertex set $V$ and edge set $E$ of a complete graph $G = (V, E)$ of order $N$, i.e. the graph with edges connecting every possible pair of vertices $v \in V$. $G$ then reflects the underlying topology of the QTPIE system in that every atom (represented by vertices) is connected to every other atom.

We now use the convenient notation $e = \overrightarrow{v_j v_i}$ for an edge $e \in E$ that connects two vertices $v_i, v_j \in V$. This can be interpreted as representing the variable $p_{ji}$, which quantifies the amount of charge transferred from atom $j$ to atom $i$. Recall that the charge-transfer variables are related to the atomic charges by Eq. (5). Using the graph-theoretic notions above, we can consider the atomic charges as a vector $\vec{q} = (q_1, \cdots, q_n) = (q_v), v \in V$ in a vector space $V_q(\mathbb{R}^n)$, which we call the atom space. Similarly, the charge-transfer variables can be used to define a vector $\vec{p} = (p_{21}, \cdots, p_{n,n-1}) = (p_e), e \in E$ in its corresponding vector space $V_p\left(\mathbb{R}^{\frac{1}{2}n(n-1)}\right)$, which we call the bond space. For example, Figure 2 shows a schematic for visualizing the elements of the atom and bond spaces for a single water molecule.

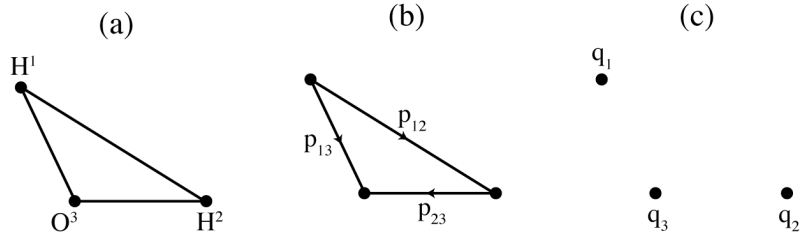

Figure 2. A schematic diagram of (a) atoms and atom pairs in a water molecule, with atoms enumerated in superscripts; (b) charge-transfer variables in bond space; and (c) charges in atom space.

We can now express the relationship between these two sets of variables in terms of a linear transformation $\mathbf{T}$ such that

$$\mathbf{T}: V_p \rightarrow V_q, \quad \mathbf{T}\vec{p} = \vec{q} \tag{13}$$

When represented by a matrix, $\mathbf{T}$ is identical to the incidence matrix of the underlying directed graph $G$, i.e. $\mathbf{T}$ is the mapping $E(G) \rightarrow V(G)$ from edges to vertices of the graph $G$, with elements $T_{ve}$ equal to 1 if the edge $e$ connects $v$ and points toward $v$, -1 if the edge $e$ connects $v$ and points away from $v$, and 0 if the edge $e$ does not connect the vertex $v$ to another vertex.

For QTPIE, QEq and similar models, the lack of topological constraints on the flow of charge means that the underlying graph $G$ must be the complete graph of order $N$, which is the graph where each vertex is connected to every other vertex. Furthermore, the antisymmetry of the charge-transfer variables specifies a particular orientation of the graph. We therefore obtain the incidence matrix for all such models as:

$$T_{ve} = \delta_{va} - \delta_{vb}, e = \overrightarrow{ba} \tag{14}$$

It is easy to verify that Eqs. (13) and (14) recover Eq. (5). We now wish to compute a transformation matrix $\mathbf{T}'$ that serves as a suitable transformation in the reverse sense, i.e.

$$\mathbf{T'} : V_q \to V_p, \quad \mathbf{T'}\vec{q} = \vec{p} \tag{15}$$

Since $V_q$ and $V_p$ in general have different dimensions, $\mathbf{T}$ is a rectangular matrix. Thus $\mathbf{T'}$ in general cannot be a true inverse, but must be the pseudoinverse, or generalized inverse, that satisfies the Moore-Penrose conditions.[32]

It is straightforward to verify that the elements of $\mathbf{T'}$ are simply

$$\left(\mathbf{T'}\right)_{ev} = \frac{\delta_{va} - \delta_{vb}}{N}, e = \overline{ba} \tag{16}$$

so that the inverse relation between the charge and charge-transfer variables is simply

$$p_e = \sum_{v \in V} \mathbf{T'}_{ev} q_v = \sum_{v \in V} \frac{\delta_{va} - \delta_{vb}}{N} q_v = \frac{q_a - q_b}{N} \quad \forall e = \overline{ba} \in E, \quad b, a \in V \tag{17}$$

This relation has a completely unexpected simplicity that allows the original QTPIE energy function of Eq. (1) to be rewritten as

$$E = \sum_i q_i \sum_j \frac{f_{ji} - f_{ij}}{N} + \tfrac{1}{2} \sum_{ij} q_i q_j J_{ij} \tag{18}$$

This is our main result, which gives a much more compact set of working equations as we can now solve the QTPIE model in exactly the same way as the QEq model. This reformulated problem is mathematically equivalent to Eq. (4) in that the predicted charge distributions are identical. However, the reformulated problem given in Eq. (18) is much more amenable to solution with conventional linear algebra algorithms as we have analytically projected out the nullspace in the construction of $\mathbf{T'}$.

This reformulation in Eq. (18) of the QTPIE model is more than just a mathematical convenience, as it also furnishes some insight into why the model works as well as it does. By substituting our definition of the pairwise electronegativity[29] $f_{ij} = \chi_j k_{ij} S_{ij}$ into (1), where $\chi_j$ is the electronegativity of atom $j$, $k_{ij} = k_{ji}$ is a symmetric constant factor, and $S_{ij}$ is the overlap integral between atoms $i$ and $j$, we obtain

$$E = \sum_i q_i \sum_j k_{ij} \frac{(\chi_i - \chi_j) S_{ij}}{N} + \frac{1}{2} \sum_{ij} q_i q_j J_{ij} \qquad (19)$$

Interestingly, the effect of introducing the bond pairwise electronegativity is to renormalize the atomic electronegativities by an amount that depends on the electronegativities of all other atoms in the system. We previously introduced the overlap integrals as strongly distance-dependent attenuation factors that would allow the charge model to exhibit the correct asymptotic behavior at dissociation limits,[27,29] as required by the appearance of derivative discontinuities in density functional theory.[26,27,31,60-63] These overlap integrals now appear in the atom-space formulation as weighting factors for the averaging of electronegativity differences.

We also note that the symmetric factor $k_{ij}$ must scale as $N$ in order to guarantee the correct size-extensivity of the atomic electronegativities in QTPIE. Therefore, we define

$$k_{ij}^{-1} = \sum_{j'} S_{ij'} / N \qquad (20)$$

independent of the index $j$. This leads to the energy function

$$E = \sum_i q_i \frac{\sum_j (\chi_i - \chi_j) S_{ij}}{\sum_{j'} S_{ij'}} + \frac{1}{2} \sum_{ij} q_i q_j J_{ij} \qquad (21)$$

Defining the effective atomic electronegativity $v_i$ to be the coefficient of the linear term above and introducing the chemical potential $\mu$ to enforce the charge conservation constraint $\sum_i q_i = 0$, QTPIE in atom space reduces to solving the linear system

$$\begin{pmatrix} \mathbf{J} & \mathbf{1} \\ \mathbf{1}^T & 0 \end{pmatrix} \begin{pmatrix} \mathbf{q} \\ \mu \end{pmatrix} = \begin{pmatrix} -\mathbf{v} \\ 0 \end{pmatrix} \qquad (22)$$

The matrix above is real, symmetric and full-rank, but indefinite, necessitating some care in the numerical algorithms used to solve Eq. (22).

## Results and Discussion

The preceding analysis shows that there a mathematical equivalence between models formulated using either atomic charges (atom space) or charge-transfer variables (bond space). QTPIE was simpler to formulate in bond space as this allowed us to construct electronegativities that are explicitly pairwise dependent. However, Eq. (22) has significantly lower computational complexity owing to the smaller size of the linear system in atom space and thus has a significant numerical advantage over the corresponding bond-space formulation of Eq. (4). The pair of transformations given by Eqs. (5) and (17) thus allows us to have the best of both spaces.

We now turn to the pragmatic issue of solving Eqs. (4) and (22). Figure 3 shows how the execution time of the various implementations of the QTPIE model scales with system size for linear water chains of increasing length (one representative water chain is shown in the inset). We do not exploit sparsity in any way for these tests, using dense matrix multiplication and factorization routines throughout. Thus, these results should be considered as upper bounds on the computational complexity of the various solution methods.

We solved the model in the bond space using both SVD and COD approaches detailed above. Singular value decomposition was carried out using the DGELSS routine from the LAPACK linear algebra library.[59] The COD method was implemented using routines from the LAPACK and BLAS libraries. The algorithm is similar to the LAPACK routine DGELSY, but without time-consuming step of numerical rank determination since for our problems the rank of these matrices are known exactly. Both SVD and COD methods scale as $O(N^6)$ for the range of system sizes investigated here. However, the COD method is faster by roughly an order of magnitude. In practice, we also find that COD tended to be numerically unstable without preconditioning; however, a simple diagonal (Jacobi) preconditioner was sufficient to observe convergence.

In contrast, the reformulated atom-space model of Eq. (22) is much more efficient due to the intrinsically smaller matrix sizes. A direct solution using the DGESV routine from LAPACK showed an asymptotic complexity of $O(N^{2.6})$ while an implementation of the iterative generalized minimal residuals (GMRES) algorithm[64] using dense matrix multiplications exhibited an asymptotic complexity of $O(N^{1.8})$. The charges computed using both atom-space methods are identical essentially to within machine precision, while the charges computed using the bond-space methods agree to three decimal places, reflecting the greater intrinsic numerical instability of the bond-space problem. Thus as expected, our model is much more practical to solve when reformulated in atom-space charge variables compared to its original formulation in bond-space charge-transfer variables.

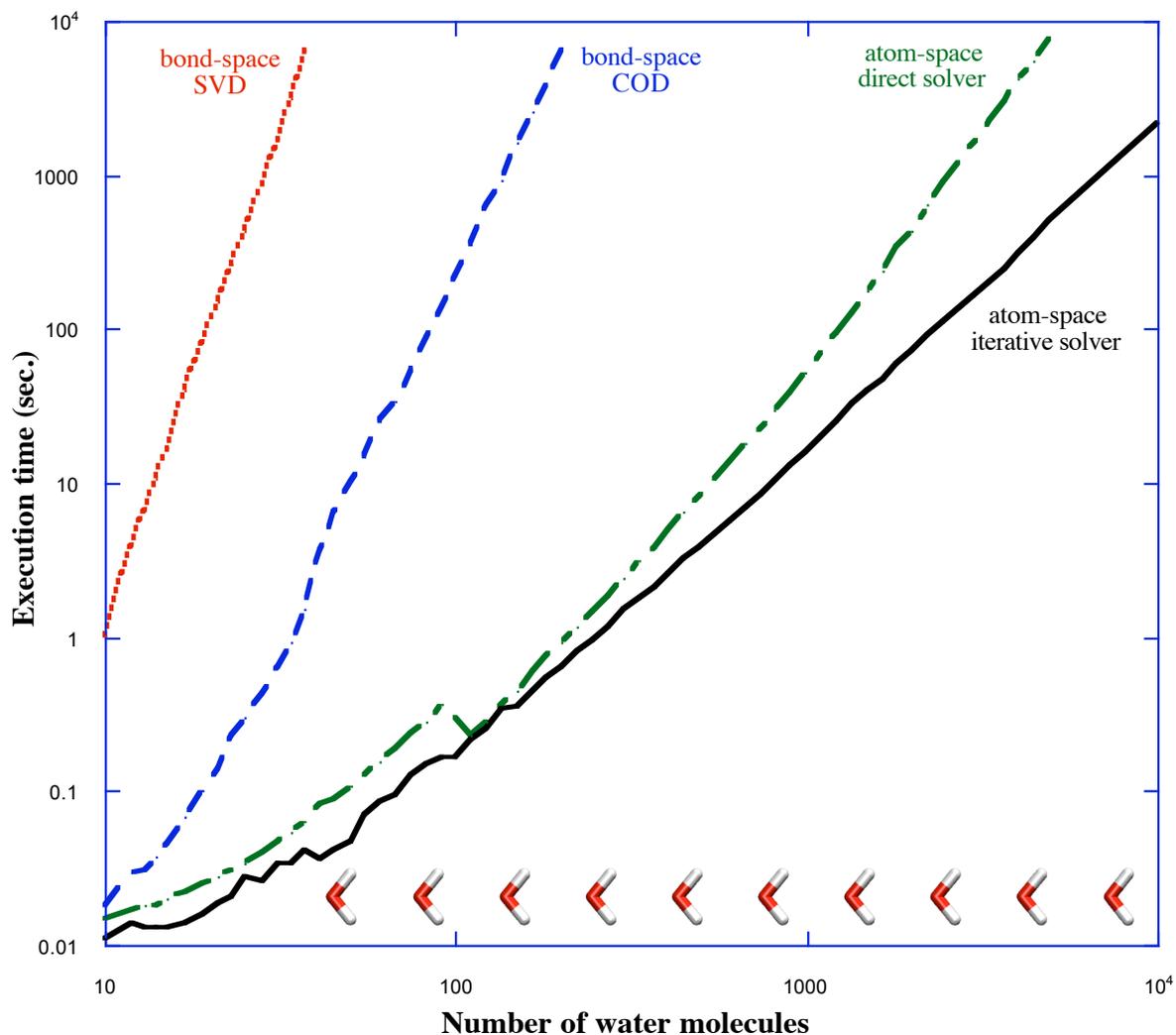

Figure 3. Execution time of QTPIE for coplanar, linear chains of water molecules for four methods of solving the system of equations: bond-space singular value decomposition (SVD) (red dotted line), bond-space complete orthogonal decomposition (COD) (blue dashed line), atom-space direct matrix solver (green dash-dotted line), and atom-space iterative solution using the generalized minimal residuals (GMRES) algorithm (black solid line). All calculations were run on a single core of an AMD Opteron 175 CPU with 2.2 GHz clock rate.

The transformations of Eqs. (5) and (17) illustrate the existence of an underlying topological duality between models formulated in atom space, such as QEq, and models formulated in bond space, such as QTPIE. With these transformations, any bond-space model can be related to an equivalent atom-space model that predicts the same charge distribution, and vice versa. The reformulation of atom-space models in bond space is trivial. For the reverse case, consider the most general charge model in bond space that has a quadratic energy function:

$$E = \sum_{i,j=1}^{N} L_{ij} p_{ji} + \sum_{i,j,k=1}^{N} M_{ijkl} p_{ji} p_{lk} \tag{23}$$

It is straightforward to show from Eq. (17) that this bond-space model is exactly equivalent to the analogous model in atom space

$$E = \sum_{i=1}^{N} X_i q_i + \sum_{i,j=1}^{N} Y_{ij} q_i q_j \tag{24}$$

where

$$X_i = \sum_{j=1}^{N} \left( L_{ij} - L_{ji} \right) / N \tag{25}$$

$$Y_{ij} = \sum_{k,l=1}^{N} \left( M_{ikjl} - M_{ilkj} - M_{kijl} + M_{kilj} \right) / N^2 \tag{26}$$

Thus any quadratic charge model in bond space can be rewritten exactly as an equivalent quadratic charge model in atom space, which can be solved more efficiently. This analysis can be easily extended to much more general charge models containing terms of arbitrary order in the charge-transfer variables and atomic charges respectively.

Our analysis also provides a straightforward prescription for deriving mappings analogous to Eqs. (5) and (17) for fluctuating-charge models with *a priori* topological constraints on charge flow.[3,9,10,12-16,20,48-51,65,66] In such models, the mapping from atom space to bond space is still represented by the incidence matrix of the graph encoding the topological constraints. Although the reverse mapping **T′** may not be as simple as Eq. (17), it can nevertheless be computed using any method for calculating the pseudoinverse. Interestingly, there exists a specialized algorithm for calculating the pseudoinverse of arbitrary incidence matrices.[67] At any rate, the transformation need only be calculated once for any given model — it is only necessary to recompute $\mathbf{T}'$ when the incidence matrix changes. Furthermore, as long as both mappings have rank $N - 1$, the bijection between bond-space models and atom-space models will hold.

## Electrical circuit representations of fluctuating-charge models

In order to gain further physical insight into fluctuating-charge models, we investigate the physical systems that are described by the same working equations. It turns out that dc circuits of capacitors and batteries can be described with the same energy functions as fluctuating-charge models. In other words, atomic systems are described by dc electrical circuits in fluctuating-charge models.

Fluctuating-charge models assume thermodynamic equilibrium, and therefore the resulting charge distributions they predict must be stable to fluctuations in time. Thus, they can only describe dc circuits in the absence of any net current flow. Therefore, electronegativity equilibration models can only correspond to circuits that contain capacitors and batteries, i.e. dc sources of electromotive forces, since these are the only elementary electrical circuit components that exhibit nontrivial behavior in the absence of any net electrical current flow.

To illustrate the concepts that we will use later, we will consider first the very simple circuit of Figure 4, consisting of a single battery with electromotive force (colloquially termed 'voltage') $V$ connected to a lone capacitor of capacitance $C$.

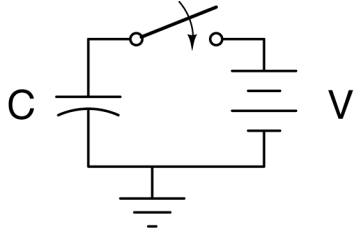

Figure 4. A minimal circuit with one capacitor and one battery. Each atomic site in QEq can be represented as such a minimal circuit element.

We now calculate the charge $q$ that eventually accumulates on the capacitor once the circuit is closed using a Hamiltonian approach. The energy function for the circuit in Figure 4 is

$$E(q) = -Vq + \tfrac{1}{2}C^{-1}q^2 \tag{27}$$

where the first term is the work done by the system to charge up the capacitor and is subtracted from the total energy. For this circuit to be in equilibrium the total energy must be minimized; elementary calculus then derives the constitutive relation $q = CV$. The enforcement of the total charge on a system can be implemented straightforwardly by introducing a Lagrange multiplier $\mu$ that corresponds to the chemical potential. This effectively shifts the bias of the ground to a non-zero voltage $\mu$.

The preceding discussion can be used in principle to relate any fluctuating-charge model with the existence of batteries and capacitors respectively. We note that others have explored similar ideas in defining connections between fluctuating charge models and electrical circuit theory, but using resistors instead.[3,48] Existing circuit duality identities permit the transformation of our capacitor-battery circuits into current-resistor circuits; however, we believe that the capacitor-battery circuit model is physically more reasonable since in the absence of magnetic fields, it is not possible for classical physical systems to sustain quasi-steady currents.

*A circuit representation of QEq*

We now show that the QEq model corresponds to the circuit in Figure 5, created by making $N$ copies of the minimal circuit in Figure 4 and connecting them all together with a common ground with bias $\mu$. The dashed lines along the wires denote multiple copies of the minimal circuit omitted from the diagram, while the additional dotted lines between the capacitors represent additional terms arising from mutual interactions between the charge distributions of each capacitor.

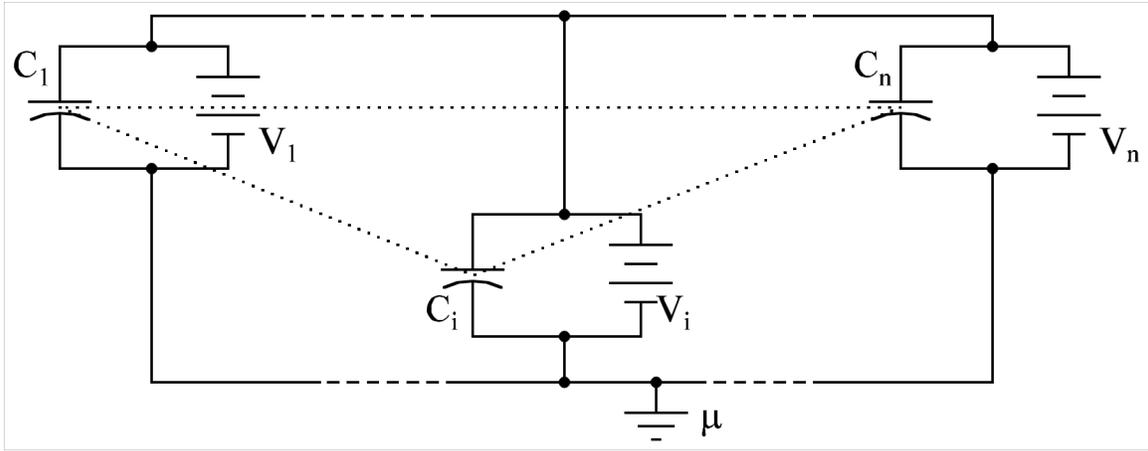

Figure 5. The circuit representing QEq, consisting of $n$ minimal circuits (atoms) connected by a common ground.

The energy function corresponding to the circuit in Figure 5 is

$$E\left(q_1,\cdots,q_i,\cdots,q_n\right) = \sum_{i=1}^{N}\left(-V_i q_i + \tfrac{1}{2}C_i^{-1}q_i^2\right) + \sum_{i<j}\left(C^{-1}\right)_{ij}q_i q_j \qquad (28)$$

where the extra terms are parameterized in terms of coefficients of induction[68] $\left(C^{-1}\right)_{ij}$ that represent the mutual Coulomb interactions of the charges built up on every capacitor. Again, we introduce a bias $\mu$ to the ground voltage as a Lagrange multiplier to enforce the constraint on the total charge $Q = \sum_{i=1}^{N} q_i$. In comparison, the QEq model[7] for a $N$-atom system has the form:

$$E(q_1, \cdots, q_n) = \sum_{i=1}^{N} \left( \chi_i q_i + \tfrac{1}{2} \eta_i q_i^2 \right) + \sum_{i<j} J_{ij} q_i q_j \qquad (29)$$

The QEq parameters map perfectly onto the parameters describing a capacitor-battery pair: the electronegativities $\chi_i = -V_i$ are directly related to internal electromotive forces, the chemical hardnesses $\eta_i = C_i^{-1}$ are identical to elastances or inverse capacitances, and the screened Coulomb interactions $J_{ij} = \left( C^{-1} \right)_{ij}$ are equivalent to coefficients of inductance[68]. Furthermore, these relations are dimensionally consistent. Therefore, with only a minor relabeling of the relevant quantities, the QEq model is equivalent to the electrical circuit in Figure 5.

*A circuit representation of QTPIE*

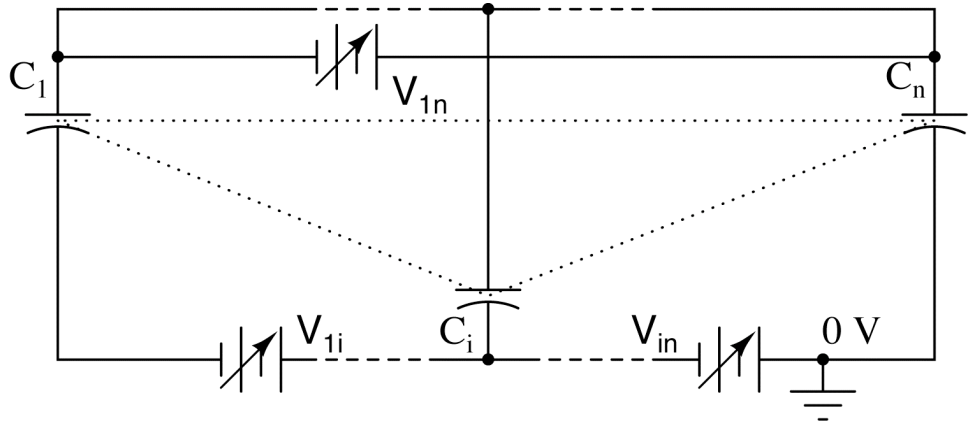

Figure 6. The circuit diagram corresponding to the QTPIE charge model.

From a similar argument we can construct a circuit diagram corresponding to the QTPIE model, shown in Figure 6.[29] (Wires not meeting at a dot junction are not connected.) In the same way we constructed the QTPIE model from QEq, we obtain this circuit diagram in two steps. First, the transformation of variables is equivalent to replacing all batteries in Figure 5 with equivalent batteries connected along all possible pairs of capacitors. Second, to obtain the QTPIE model, the only essential modification of the QEq model was to allow the electronegativities to be pairwise dependent on the distance between pairs of atoms. Hence, the corresponding circuit elements must be variable voltage dc sources straddling each pair of capacitors.

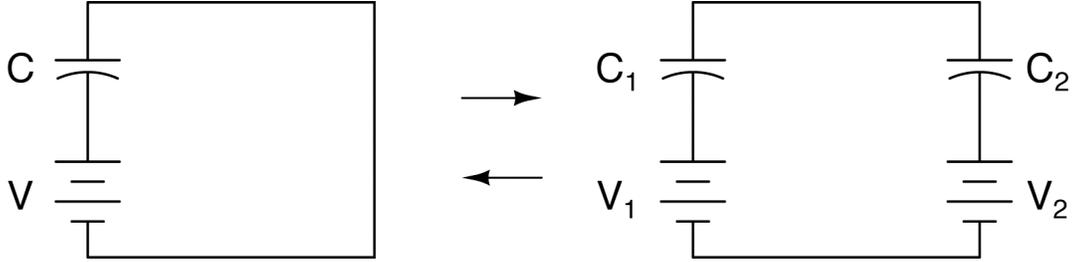

Figure 7. An illustration of the circuit diagrams for a bond-space fluctuating-charge model (left) and an atom-space fluctuating-charge model (right) for a diatomic system.

The relationship between bond-space and atom-space fluctuating charge models is intimately related to the notion of circuit duality.[69] We can illustrate this for a diatomic system, whose corresponding circuits in bond space and atom space are given in Figure 7. The bond-space equation is given simply by

$$C^{-1}q = V \tag{30}$$

and the atom-space equation is

$$\begin{pmatrix} C_1^{-1} & J \\ J & C_2^{-1} \end{pmatrix} \begin{pmatrix} q_1 \\ q_2 \end{pmatrix} = \begin{pmatrix} V_1 \\ V_2 \end{pmatrix} \tag{31}$$

where charge conservation requires $q_1 = -q_2$. Topological considerations give the incidence matrix for this system as $\begin{pmatrix} 1 & -1 \end{pmatrix}^T$, which has pseudoinverse $\frac{1}{2}\begin{pmatrix} 1 & -1 \end{pmatrix}$. To transform Eq. (30) into a form of the type given in Eq. (31), apply the mapping of Eq. (17) so that

$$\begin{pmatrix} 1 \\ -1 \end{pmatrix} \left( C^{-1} \right) \frac{1}{2} \begin{pmatrix} 1 & -1 \end{pmatrix} \begin{pmatrix} 1 \\ -1 \end{pmatrix} (q) = \begin{pmatrix} 1 \\ -1 \end{pmatrix} V \tag{32}$$

which simplifies to

$$\frac{1}{2} \begin{pmatrix} C^{-1} & -C^{-1} \\ -C^{-1} & C^{-1} \end{pmatrix} \begin{pmatrix} q \\ -q \end{pmatrix} = \begin{pmatrix} V \\ -V \end{pmatrix} \tag{33}$$

This tells us that the dual circuit with $C_1^{-1} = C_2^{-1} = -J = \frac{1}{2}C^{-1}$ and $V_1 = -V_2 = V$ predicts a charge distribution $q_1 = -q_2 = q$ that is compatible with the bond-space charge transfer under the mapping of Eq. (17).

Conversely, to transform Eq. (31) into a form similar to Eq. (30), apply the mapping of Eq. (5) to obtain

$$\frac{1}{2}\begin{pmatrix} 1 & -1 \end{pmatrix}\begin{pmatrix} C_1^{-1} & J \\ J & C_2^{-1} \end{pmatrix}\begin{pmatrix} 1 \\ -1 \end{pmatrix}\frac{1}{2}\begin{pmatrix} 1 & -1 \end{pmatrix}\begin{pmatrix} q_1 \\ q_2 \end{pmatrix} = \frac{1}{2}\begin{pmatrix} 1 & -1 \end{pmatrix}\begin{pmatrix} V_1 \\ V_2 \end{pmatrix} \tag{34}$$

Taking into account charge neutrality, this simplifies to

$$\left( C_1^{-1} - 2J + C_2^{-1} \right)q_1 = \left( V_1 - V_2 \right) \tag{35}$$

In this example, the presence of Kirchhoff's law in the mapping of Eq. (5) allowed us to derive the well-known combination rule for capacitance $C^{-1} = C_1^{-1} + C_2^{-1}$ (by neglecting the Coulomb coupling $J$ as is common in classical circuit analysis), and the combination rule for voltages $V = V_1 - V_2$, noting that $V_2$ is the voltage of a battery oriented in opposite way relative to $V_1$. In addition, the charge distribution $q_1 = -q_2 = q$ is indeed what we had expected.

Finally, we note that while is tempting to associate the charge transfer variables with electrical currents, this is dimensionally inconsistent since they have dimensions of charge, not current. However, a consistent interpretation of these variables is that they are integrated traces of transient currents as the system equilibrates. The variables defined in this manner retain the property of detailed balance, yet are compatible with the concept of equilibrium since there are no net current flows. Although external potentials induce current flow, the buildup of charge in the capacitors decreases the potential difference driving such currents, eventually establishing equilibrium when the potentials are equalized. By identifying such potentials as electronegativities, we therefore see that in the QTPIE model, Sanderson's principle of electronegativity equalization[70] comes from the formation of countercurrents induced from polarization effects, i.e. charge buildup in atomic capacitors.

## Conclusions

Our previously introduced QTPIE model is a fluctuating-charge model that exhibits correct asymptotic behavior for dissociating molecular systems. Formulating our new model in terms of charge-transfer variables allows us to construct explicitly distance-dependent pairwise electronegativities. However, the linear system of Eq. (4) which determines the bond space variables that minimize the QTPIE energy exhibits significant linear dependencies, complicating numerical solution. We have discovered that the rank deficiency in our QTPIE model, and in bond-space fluctuating-charge models in general, can be attributed to the conservative nature of the laws of electrostatics, thus showing that the rank deficiency has a genuine physical basis, and is not merely a numerical inconvenience. With this knowledge, we constructed a numerical algorithm based on complete orthogonal decomposition that had better asymptotic complexity than singular value decomposition; this allowed an order of magnitude reduction in the time needed to solve Eq. (4). However, the computational complexity of this algorithm was still considerably higher than that for solving atom-space models.

We then showed that each fluctuating-charge model defined in bond space is equivalent *via* the mappings of Eqs. (5) and (17) to a related model of the form given in Eq. (22) formulated in atom space that predicts exactly the same charge distribution. Therefore, it is possible to formulate fluctuating-charge models with pairwise electronegativities that nonetheless retain the same asymptotic computational complexity as conventional atom-space models. In the process, we have discovered a framework which unifies models with and without topological constraints. In particular, we have shown that the underlying graphical structure of a topologically unconstrained fluctuating-charge model is that of a complete directed graph; thus fluctuating-charge models can be considered a special case of a larger class of graph charge models.

Finally, the QEq and QTPIE fluctuating-charge models can be described using the classical theory of electrical circuits, but with Coulomb interactions playing a significant role on the atomic scale that could otherwise be neglected in the description of macroscopic circuits. The circuit interpretation helps us establish some intuition for the duality mappings (5) and (17).

## Acknowledgments

This work was supported by DOE DE-FG02-05ER46260.

# Appendix to 'A unified theoretical framework for fluctuating-charge models in atom space and bond space'

Jiahao Chen and Todd J. Martínez

Department of Chemistry, Frederick Seitz Materials Research Laboratory and

The Beckman Institute

University of Illinois at Urbana Champaign

Urbana, IL 61801

# Appendix: a proof of the rank deficiency property using matroid theory

In this appendix we provide a formal proof that the linear response kernel $\mathbf{A}$ must have rank $N-1$ as stated above, which was justified from an intuitive counting argument of the degrees of the freedom in the problem. The proof is most elegantly stated in the language of matroid theory.1. We use the notation $|X|$ to denote the cardinality of a set $X$ and furthermore assume familiarity with basic concepts of set theory and graph theory. We omit proofs of established results which may be found in any standard text on matroid theory. For completeness, we provide the following definitions and results to be used later.

**Definition A1**. A matroid $M$ is an ordered pair $(E, \mathfrak{I})$ where $\mathfrak{I}$ is the set of subsets of $E$ such that $\mathfrak{I}$ contains the empty set, i.e. $\{\ \} \in \mathfrak{I}$, all subsets of elements of $\mathfrak{I}$ are themselves elements of $\mathfrak{I}$, i.e. for all $I \in \mathfrak{I}$ and $I' \subset I$, then $I' \in \mathfrak{I}$, and $\mathfrak{I}$ obeys the independence augmentation axiom, i.e. for all $I_1, I_2 \in \mathfrak{I}$ such that $|I_1| < |I_2|$, $e \in I_2 - I_1$ such that $I_1 \cup \{e\} \in \mathfrak{I}$. $E$ is called the ground set of $M$ and an element of $\mathfrak{I}$ is called an independent set.

**Definition A2**. Two matroids $M1$ and $M2$ are isomorphic, denoted $M_1 \cong M_2$, if there exists a bijection $f: E(M_1) \rightarrow E(M_2)$ between the base sets of each matroid, and any subset $X \subseteq E(M_1)$ in $M1$ is independent if and only if its image $f(X)$ is also independent in $M2$.

**Lemma A3**. Let $\mathbf{A}$ be a real square matrix of dimension $\frac{1}{2}n(n-1)$ with columns $a_i$. Then there exists a matroid $M[\mathbf{A}]$ called the vector matroid induced by $\mathbf{A}$ with columns $E = \left\{ a_v : v = 1, ..., \frac{1}{2} n(n-1) \right\}$ forming the ground set and independent sets as linearly independent subsets of $E$. $\mathbf{A}$ is called a $\mathbb{R}$-representation of $M[\mathbf{A}]$.

**Lemma A4**. Let $G = (V(G), E(G))$ be a graph with vertices $v_i \in V(G)$ and edges $e_{ij} \in E(G)$ connecting $v_i$ and $v_j$. Then there exists a matroid $M(G)$ called the cycle matroid with ground set equal to the edge set, i.e. $E(M(G)) = E(G)$ and independent sets corresponding to acyclic subsets of $E(G)$. We note that $G = K_N$, the graph of $N$ vertices connected by all possible unique edges, describes the connectivity of our QTPIE charge model. We do not assume any a priori connectivity information and hence our model must in principle consider all possible pairwise charge transfers. This is formalized in the following lemma.

**Lemma A5**. Consider the linear algebra problem $\sum_v A_{v\lambda} p_v = V_\lambda$ where $\mathbf{A}$ is a real square matrix of dimension $\frac{1}{2} N(N-1)$. Then there exist a bijection between $E(M[\mathbf{A}])$ and $\{p_v\}$, a bijection between $E(M[\mathbf{A}])$ and $\{V_v\}$, and a bijection between $E(M[\mathbf{A}])$ and $E(K_N)$.

**Proof**. The first two are trivial, since any index of the columns of $\mathbf{A}$ also indexes the corresponding row entry of $p_v$. From the rules of matrix-vector multiplication, the identity mapping

$$f : E(M[\mathbf{A}]) \to \{p_v\}, f(a_v) = p_v \tag{1}$$

is a trivial bijection. Furthermore since $\mathbf{A}$ is symmetric, every column of $\mathbf{A}$ is identical to the transpose of its corresponding row, and so the rules of matrix multiplication also show that the identity mapping

$$g : E(M[\mathbf{A}]) \to \{V_v\}, f(a_v) = V_v \tag{2}$$

is another trivial bijection. The third bijection is the following identity mapping

$$h : E\big(M[\mathbf{A}]\big) \rightarrow E\big(M(K_N)\big) = E(K_N), h(a_v) = e_{ij} \tag{3}$$

where $e_{ij}$ is the edge connecting the vertices $v_i$ and $v_j$ and $v$ is related to $i$ and $j$ by

$$v(i,j) = \tfrac{1}{2}\max(i,j)\big(\max(i,j) - 1\big) + \min(i,j) \tag{4}$$

∎

An immediate consequence of the preceding lemma is that the each edge $e_{ij} \in E(K_n)$ in the graph $K_n$ can be associated with two weights $p_v$ and $V_v$.

We now prove that Kirchhoff's voltage law determines the various properties of $\mathbf{A}$ that we had claimed earlier. To do so we first prove this main theorem.

**Theorem A6**. Let $\mathbf{A}$ be the matrix defined in the preceding lemma and furthermore let the set $\{V_v\}$ obey the holonomic constraints that for all dependent subsets $X \subseteq E(K_N)$, $\displaystyle\sum_{v \in g\left(h^{-1}(X)\right)} v = 0$. Then $M[\mathbf{A}] \cong M(K_N)$.

**Proof**. The map $h$ defined in the preceding lemma provides the necessary bijection to demonstrate isomorphism. Now consider $X \subseteq E\big(M[\mathbf{A}]\big)$. We now want to show that $h(X) \subseteq E\big(M(K_N)\big)$ is independent in $M(K_N)$ if and only if $X \subseteq E\big(M[\mathbf{A}]\big)$ is independent in $M[\mathbf{A}]$.

First suppose that $X \subseteq E\big(M[\mathbf{A}]\big)$ is a dependent set. Then its elements must be linearly dependent, i.e. there exists real coefficients $\left\{c_\mu : c_\mu \in \mathbb{R} \setminus \{0\}\right\}$ such that $\displaystyle\sum_{\mu=1}^{|X|} c_\mu x_\mu = \mathbf{0}$ for $x_\mu \in X$.

This implies that $\mathbf{0} = \sum_{\mu=1}^{|X|} c_\mu V \cdot x_\mu = \sum_{\mu=1}^{|X|} c_\mu p_\mu$. Since $\{p_\mu\}$ obey detailed balance, there must exist a charge transfer variable $r = -p_\nu$ such that $c_\nu r = \sum_{\mu \neq \nu} c_\mu p_\mu$. This is only possible if there is more

than one path connecting the vertices $v_i$ and $v_j$ where one of these paths is provided by the edge

$e = h\left(f^{-1}\left(p_\nu\right)\right)$ and at least one path defined by $h\left(f^{-1}\left(X \setminus \{e\}\right)\right)$ where $\nu = \nu(i,j)$ as defined in the preceding lemma. Hence $h(X)$ contains at least one cycle and therefore $h(X)$ is dependent in $M\left(K_N\right)$. Taking the contrapositive completes proof of the backward statement.

Now suppose that $Y \subseteq E(K_N)$ is a dependent set, i.e. is a cycle. Then the constraints on $\{V_\nu\}$

immediately give $\mathbf{0} = \sum_{\mu=1}^{|Y|} V_\mu = \sum_{\mu=1}^{|Y|} x_\mu \cdot p$, showing that the elements of $h^{-1}(Y) \subseteq E\left(M[\mathbf{A}]\right)$ are

linearly dependent. Taking the contrapositive completes proof of the forward statement.■

The isomorphism established in the preceding theorem is a very powerful one, for it allows a

collaboration of concepts in linear algebra with analogous notions in graph theory. One such

instance is in generalizing the notion of basis as follows:

**Definition A7**. The set $\mathbf{B}$ is a set of bases of a matroid $M$ if and only if $\mathbf{B}$ is not empty, and $\mathbf{B}$

satisfies the basis exchange axiom, i.e. for $B_1, B_2 \in \mathbf{B}$ and $x \in B_1 - B_2$ then there is an element such that $\left(B_1 - \{x\}\right) \cup \{y\} \in \mathbf{B}$.

It immediately follows that each element of $\mathbf{B}$ is a maximally independent sets that generalizes the concept of a complete basis that spans the range of a matrix $\mathbf{A}$, and that each element of $\mathbf{B}$

has the same cardinality. The generalization of this to graphs is as follows:

**Lemma A8**. Let $G = (V(G), E(G))$ a graph with $k$ components. Then the bases of the corresponding cycle matroid $M(G)$ are the edge sets of spanning forests of $G$, each of cardinality $|V(G)| - k$.

The rank of a matroid is defined as the cardinality of any of its basis sets. The implications for our matrix **A** immediately follow:

**Corollary A9**. The matrix **A** has rank $N - 1$.

**Proof.** The complete graph $K_N$ is connected, and therefore the matroid $M(K_N)$ has rank $|V(K_N)| - 1 = N - 1$. Since $M(K_N) \cong M[\mathbf{A}]$, $M[\mathbf{A}]$ must have the same rank as $M(K_N)$. Hence $M[\mathbf{A}]$, and **A** itself in turn, must have rank $N - 1$.